\documentclass{birkjour}
 \theoremstyle{definition}
 
 \theoremstyle{remark}

 \numberwithin{equation}{section}
 
 \usepackage{subfigure}
 
 \graphicspath{{figs/}}

\begin{document}

%
%
%
%
%
%
%
%
%

\title[Logics in fungal mycelium networks]
 {Logics in fungal mycelium networks\footnote{To be published in  special issue of \emph{Logica Universalis}  --- 
 ``Logic, Spatial Algorithms and Visual Reasoning'', edited by Andrew Schumann and Jerzy Kr\'{o}l, 2022}}


\author[Adamatzky]{Andrew Adamatzky}
\address{%
Unconventional Computing Laboratory, UWE, Bristol, UK}
\email{andrew.adamatzky@uwe.ac.uk}

\author[Ayres]{Phil Ayres}
\address{%
Centre for Information Technology and Architecture (CITA), Royal Danish Academy, Copenhagen, Denmark}
\email{phil.ayres@kglakademi.dk}

\author[Beasley]{Alexander E. Beasley}
\address{%
Centre for Engineering Research, University of Hertfordshire, UK}
\email{andrew.adamatzky@uwe.ac.uk}

\author[Roberts]{Nic Roberts}
\address{%
Unconventional Computing Laboratory, UWE, Bristol, UK}
\email{andrew.adamatzky@uwe.ac.uk}

\author[Tegelaar]{Martin Tegelaar}
\address{%
Microbiology, Department of Biology, University of Utrecht, Utrecht, The Netherlands}
\email{m.tegelaar@uu.nl}

\author[Tsompanas]{Michail-Antisthenis Tsompanas}
\address{%
Unconventional Computing Laboratory, UWE, Bristol, UK}
\email{andrew.adamatzky@uwe.ac.uk}

\author[W\"{o}sten]{Han A. B. W\"{o}sten}
\address{%
Microbiology, Department of Biology, University of Utrecht, Utrecht, The Netherlands}
\email{h.a.b.wosten@uu.nl}



\date{}

\begin{abstract}
The living mycelium networks are capable of efficient sensorial fusion over very large areas and distributed decision making. The information processing in the mycelium networks is implemented via propagation of electrical and chemical signals en pair with morphological changes in the mycelium structure. These information processing mechanisms are manifested in experimental laboratory findings that show that the mycelium networks exhibit  rich dynamics of neuron-like spiking behaviour and
a wide range of non-linear electrical properties.  On an example of a single real colony of \emph{Aspergillus niger}, we demonstrate that the non-linear transformation of electrical signals and trains of extracellular voltage spikes can be used to implement logical gates and circuits. The approaches adopted include numerical modelling of excitation propagation on the mycelium network, representation of the mycelium network as a resistive and capacitive (RC) network and an experimental laboratory study on mining logical circuits in mycelium bound composites.  
\end{abstract}

\maketitle
\section{Introduction}

The fungi is among the largest, most widely distributed group of living organisms~\cite{carlile2001fungi}. Fungi can grow as individual cells or in a interconnected network of hyphae. These hyphae that grow at their tips and branch sub-apically can be compartmentalized by porous septa that can be either in a closed or open state. In the open state, cytoplasm can stream from one compartment to the other or even from hypha to hypha. In the closed state, the compartments can act as individual entities although there is still interaction with neighboring cells~\cite{wosten2013heterogeneity}. Mycelia can be thousands of years old and can cover large surface areas. The largest known mycelium, belonging to the  \emph{Armillaria genus} covers an area of 965 hectares~\cite{smith1992fungus}. The fungi show a high degree of adaptability to environmental conditions.  They are demonstrated to efficiently explore confined spaces with their hyphae \cite{hanson2006fungi, held2008examining, held2009fungal, held2010microfluidics,held2011probing}. In fact, they even form different types of hyphae within the mycelium~\cite{tegelaar2020subpopulations}.  Optimisation of the mycelial network~\cite{boddy2009saprotrophic} is quite similar to that of slime mould \emph{P. polycephalum}, e.g. in terms of proximity graphs \cite{adamatzky2009developing} and transport networks~\cite{adamatzky2012bioevaluation}. Taking into account the ubiquity, range of length scales and spatio-temporal dynamics exhibited by the fungi, they represent a promising research target within the context of unconventional computing.

The motivation of this paper is to contribute to uncovering basic mechanisms of decision making in the fungal network in terms of Boolean gates and circuits. Mechanisms of computation discovered in mycelium networks could be utilised in future designs of electrical analog computing circuits~\cite{ulmann2020analog} and to design and program computing schemes embedded into living fungal architectures~\cite{adamatzky2019fungal}.

A first step toward discovering the computing potential of fungal networks would be to estimate frequencies of logical gates and simple circuits realisable in a single fungal colony. We implement this idea using three techniques: numerical modelling of spiking events on fungal colony, modelling the colony as a resistive and capacitive (RC) network and mining logical circuits. 

\begin{figure}[!tbp]
    \centering
    \includegraphics[width=0.99\textwidth]{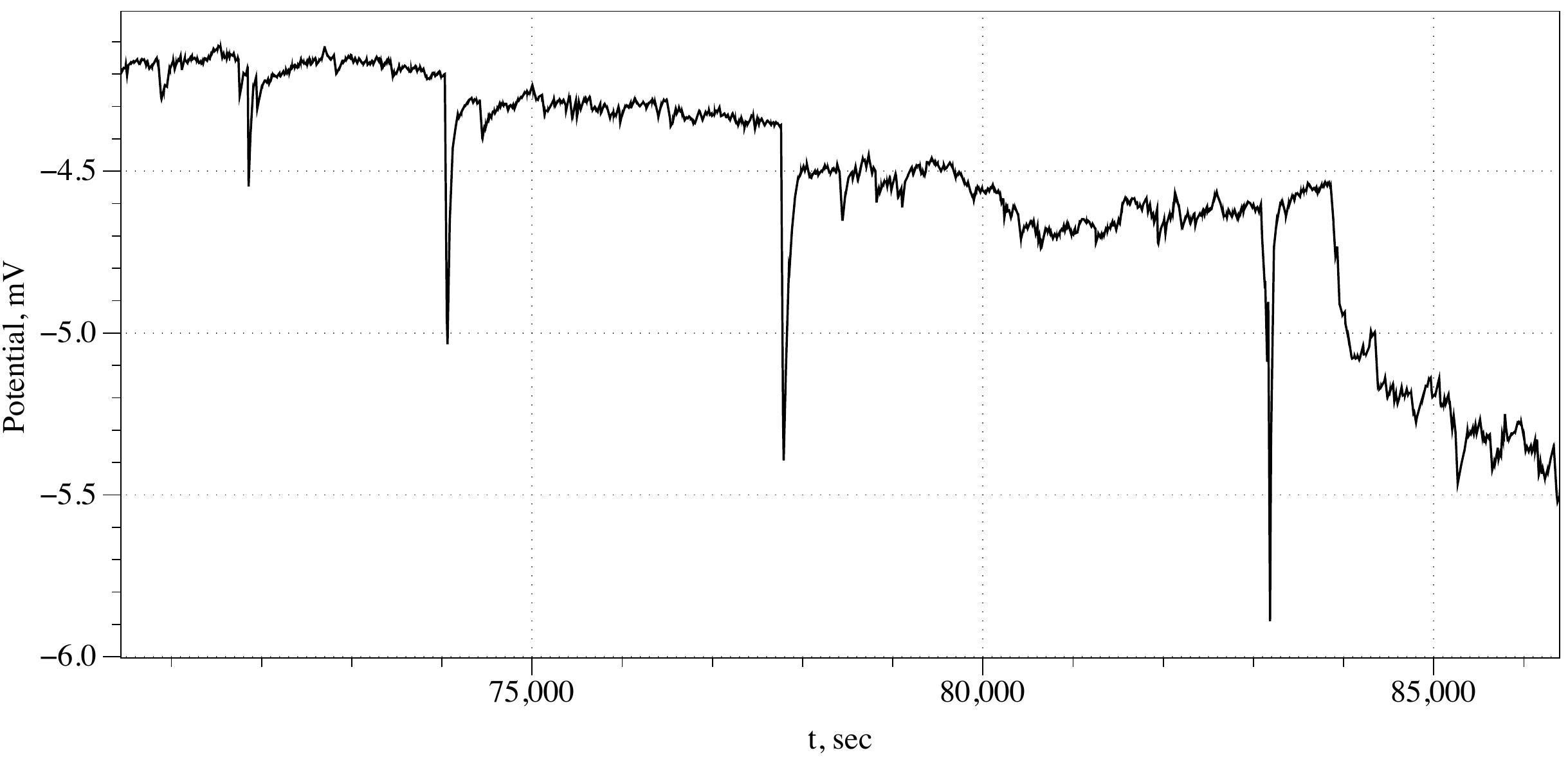}
    \caption{Exemplar spikes of extracellular electrical potential propagating in fungal mycelium.}
    \label{fig:expemplarSpikes} 
\end{figure}

In the first technique, logical gates are calculated based on the temporal co-occurrence of spikes emerging as responses to different input strings~\cite{adamatzky2020boolean}. Why do we consider spikes of electrical potential? Because these spikes are manifestations of the calcium waves that travels along mycelium networks and implement information between distant parts of the mycelium network and, possibly, participate in the information processing. First discovery of the electrical potential spikes has been done via  intra-cellular recording of mycelium of \emph{Neurospora crassa}~\cite{slayman1976action}. Further confirmed in intra-cellular recordings of action potential in hypha of \emph{Pleurotus ostreatus} and \emph{Armillaria bulbosa}~\cite{olsson1995action}, and observed in the extra-cellular recordings of fruit bodies resulting from substrates colonized by the mycelium of \emph{Pleurotus ostreatus}~\cite{adamatzky2018spiking} (Fig.~\ref{fig:expemplarSpikes}).

Spikes of fungal electrical potential are notoriously slow, with a minimum spike duration of 2 mins and maximum up to an hour. Thus the techniques of spikes based logical circuits might not be suitable for practical applications. Two other techniques exploit principles of electrical analog computing~\cite{beasley2021electrical,roberts2021mining}.
{\sc True} and {\sc False} values are represented by above threshold and below threshold voltages. Due to the non-linearity of the conductive substrate along electrical current pathways between input and output electrodes, the input voltages are transformed and thus logical mappings are implemented. 

Detailed descriptions of these techniques can be found in \cite{adamatzky2020boolean,beasley2021electrical,roberts2021mining}. Here we provide an updated overview of the approaches and provide an integrative analysis of the results.

\section{Methods}

\subsection{Colony imaging}

We have grown \emph{Aspergillus niger} fungus strain  AR9\#2~\cite{vinck2011heterogenic}. This strain expresses Green Fluorescent Protein (GFP) from the glucoamylase (\emph{glaA}) promoter. A fluorescence of GFP was localised in micro-colonies using a DMI 6000 CS AFC confocal microscope (Leica, Mannheim, Germany). Micro-colonies were imaged at 20$\times$ magnification (HC PL FLUOTAR L 20 $\times$ 0.40 DRY). Z-stacks of imaged micro-colonies were made using 100 slices with a slice thickness of 8.35~µm. 3D projections were made with Fiji~\cite{schindelin2012fiji}.

\subsection{Numerical modelling}

\begin{figure}[!tbp]
    \centering
      \includegraphics[width=0.7\textwidth]{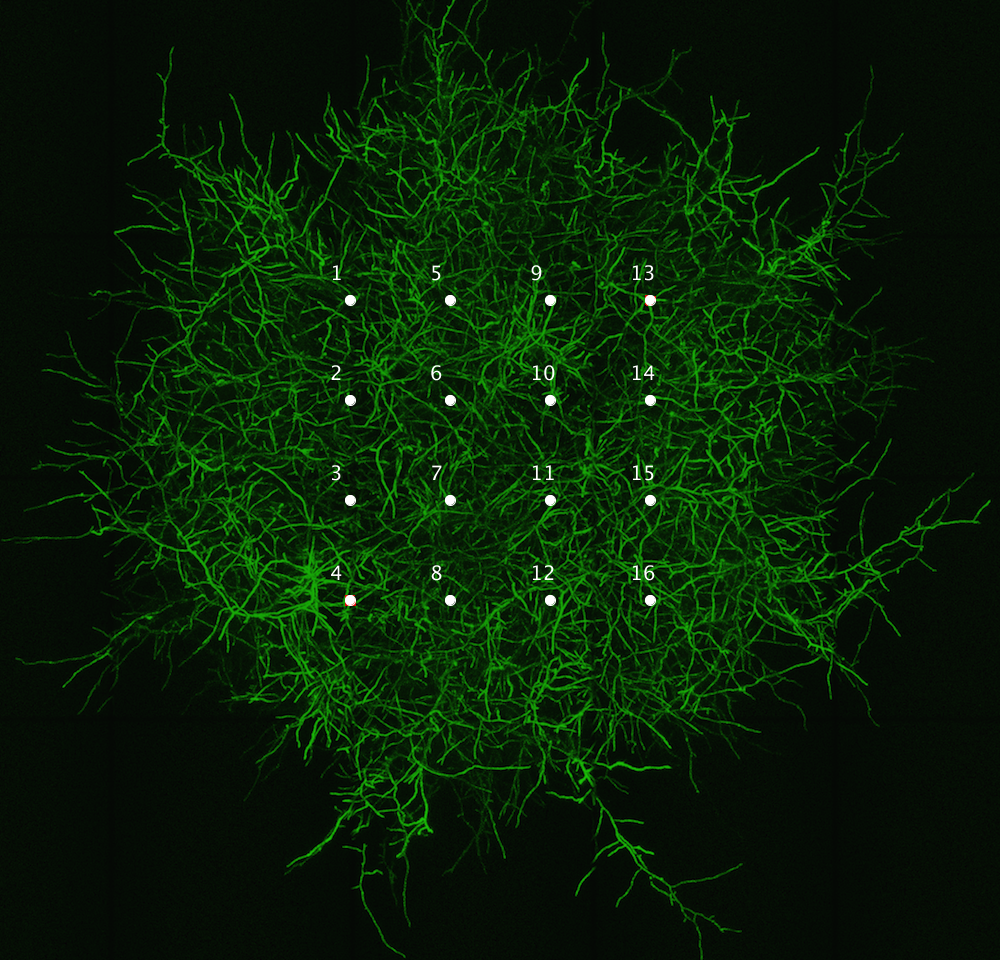}
    \caption{Image of the fungal colony, $1000 \times 960$ pixels used as a template conductive for FHN. A configuration of electrodes is superimposed on the image.}
    \label{fig:mycelium}
\end{figure}

We used a selected image of the colony, from the middle of the $z$-stack, as a conductive template. The image of the fungal colony  (Fig.~\ref{fig:mycelium}) was projected onto a $1000 \times 960$ nodes grid $C$.

We simulated electrical activity of the colony with FitzHugh-Nagumo (FHN) equations~\cite{fitzhugh1961impulses,nagumo1962active,pertsov1993spiral}:
\begin{eqnarray}
\frac{\partial v}{\partial t} & = & c_1 u (u-a) (1-u) - c_2 u v + I + D_u \nabla^2 \\
\frac{\partial v}{\partial t} & = & b (u - v),
\end{eqnarray}
where $u$ is a value of a trans-membrane potential, $v$ a variable accountable for a total slow ionic current, or a recovery variable responsible for a slow negative feedback, $I$ {is} a value of an external stimulation current. The current through intra-cellular spaces is approximated by
$D_u \nabla^2$, where $D_u$ is a conductance. 
We integrated the system using the Euler method with the five-node Laplace operator, a time step $\Delta t=0.015$ and a grid point spacing $\Delta x = 2$, while other parameters were $D_u=1$, $a=0.13$, $b=0.013$, $c_1=0.26$, $c_2=0.095$.   To show dynamics of excitation in the network, we simulated electrodes by calculating a potential $p^t_x$ at an electrode location $x$ as $p_x = \sum_{y: |x-y|<2} (u_x - v_x)$.

\subsection{Resistive and capacitive network modelling}

\begin{figure}[!tbp]
    \centering
    \includegraphics[width=0.7\textwidth]{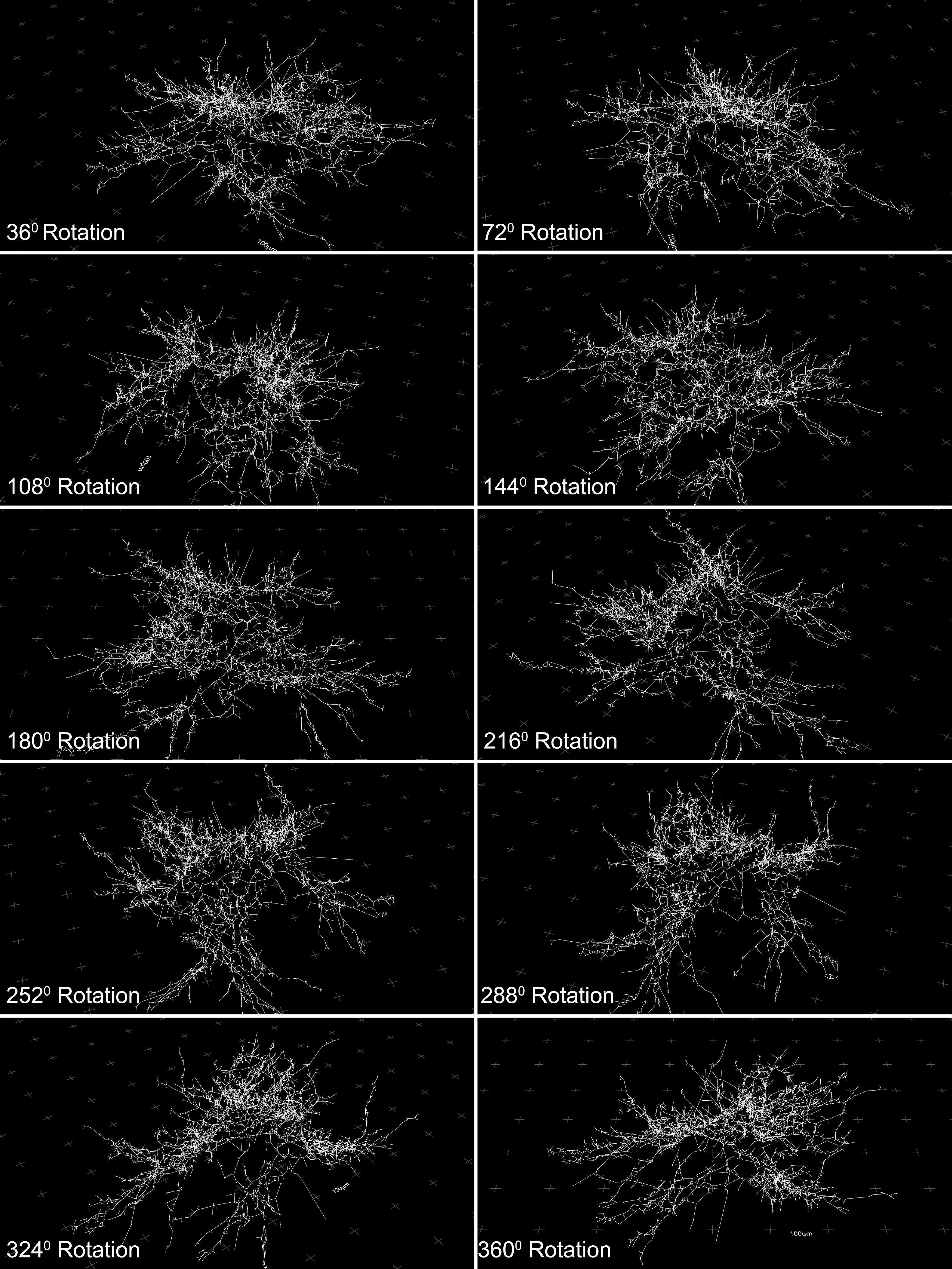}
    \caption{A graph representation of a single fungal colony.
   Each frame shows the graph after a 36$^{\circ}$ rotation around the $z$-axis.}
    \label{fig:graph}
\end{figure}

The $z$-stacks of the colony were converted to a 3D graph (Fig.~\ref{fig:graph}). The 3D graph was converted to a resistive and capacitive (RC) network, by assigning to each edge a function of a resistor or a capacitor at random. The magnitudes of the resistance and capacitance were functions of the length of the edges/connections. We have chosen resistances in the order of kOhms and values of capacitance in the order of pF. We selected ground nodes and sources (positive voltage nodes) at random. The trials were run on 1000 networks (with the same architecture but different values of resistance and capacitance). During SPICE modelling we used two voltage pulses of 60~mV on randomly chosen positive nodes.

We modelled the fungal colony in serial RC networks (resistors and capacitors are connected in series) and parallel RC networks (resistors and capacitors are connected in parallel), see basics in~\cite{horowitz1989art}. The output voltages have been binarised with the threshold $\theta$: $V>\theta$ symbolises logical {\sc True} otherwise {\sc False}.

\subsection{Experimental laboratory mining of circuits}

A hemp shavings substrate was colonised by the mycelium of the grey oyster fungi, \emph{P. ostreatus} (Ann Miller's Speciality Mushrooms Ltd, UK). Hardware was developed that was capable of sending sequences of 4 bit strings to a mycelium substrate. The strings were encoded as step voltage inputs where -5~V denoted a logical 0 and 5~V a logical 1. The hardware was based around an Arduino Mega 2560 (Elegoo, China) and a series of programmable signal generators, AD9833 (Analog, USA). 
The 4 input electrodes were 1~mm diameter platinum rods inserted to a depth of 50~mm in the substrate in a straight line with a separation of 20~mm. Data acquisition (DAQ) probes were placed in a parallel line 50~mm away separated by 10~mm. The electron sink and source was placed 50~mm on from DAQ probes. There were 7 DAQ differential inputs from the mycelium substrate to a Pico 24 (Pico Technology, UK) analogue-to-digital converter (ADC), the 8th channel was used to pass a pulse to the ADC on every input state change. There were a total of 14 repeats.
A sequence of 4 bit strings counting up from binary \textit{0000} to \textit{1111}, with a state change every hour, were passed into the substrate. Boolean strings were extracted from the data, where a logic ‘1’ was noted for a channel if it had a peak outside the threshold band for a particular state, else a value of ‘0’ was recorded. The polarity of the peak was not considered. The sum of products (SOP) Boolean functions were calculated for each output channel. For each repeat there were 7 channels and 32 thresholds giving total of 3136 individual truth tables. 

\section{Results}

\subsection{Spikes derived logical gates}

\begin{figure}[!tbp]
    \centering
    \includegraphics[width=0.9\textwidth]{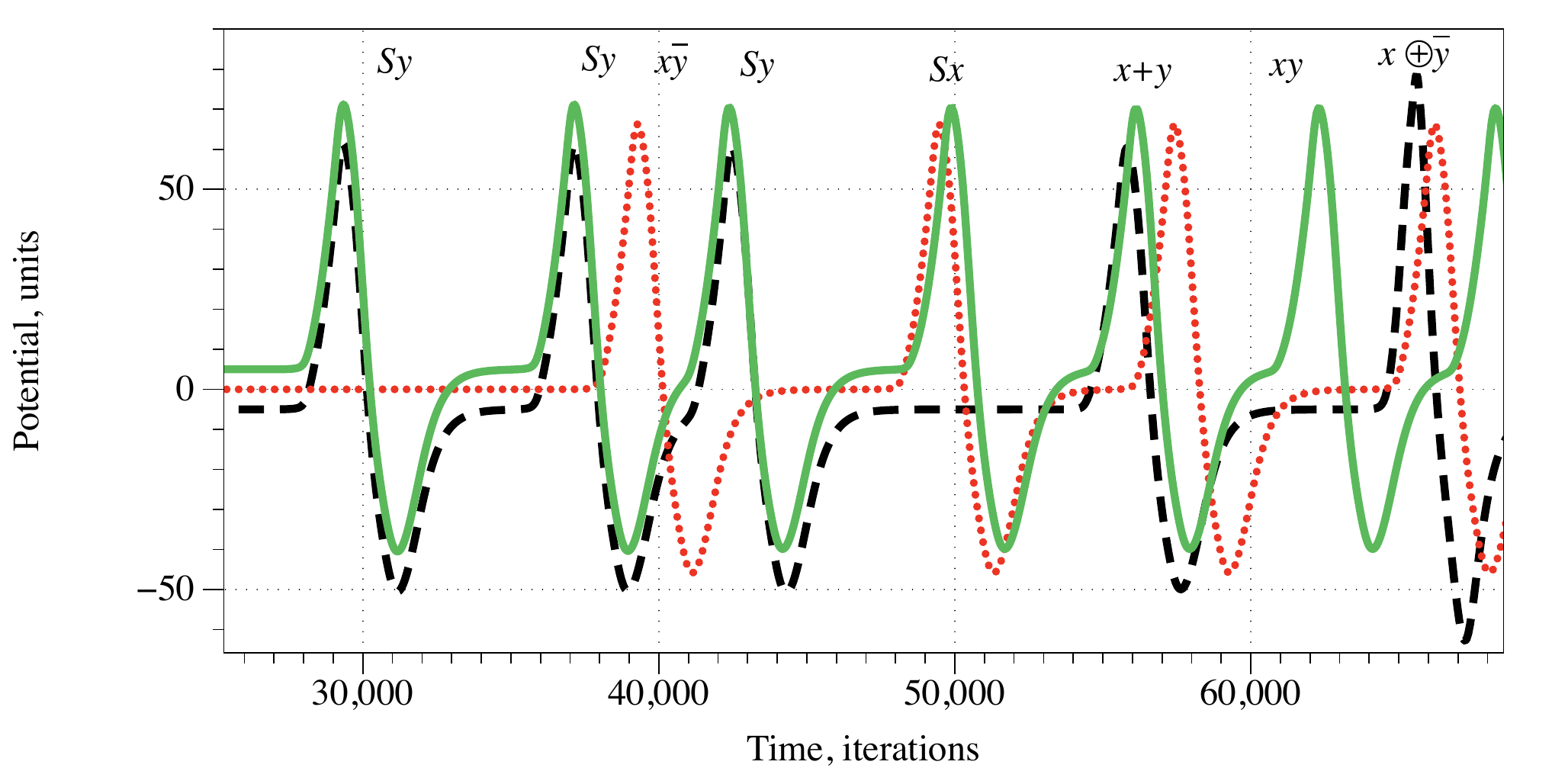}
      \caption{Examples of electrical potential spikes recorded on the electrode 7. The data represent responses to input impulse strings, entered via electrodes 5 and 15, inputs (01), black dashed line, (10), red dotted line, (11), solid green line. The locations of electrodes are shown in  Fig.~\ref{fig:mycelium}. }
    \label{fig:5_15_example}
\end{figure}

\begin{figure}[!tbp]
    \centering
  \includegraphics[height=0.9\textheight]{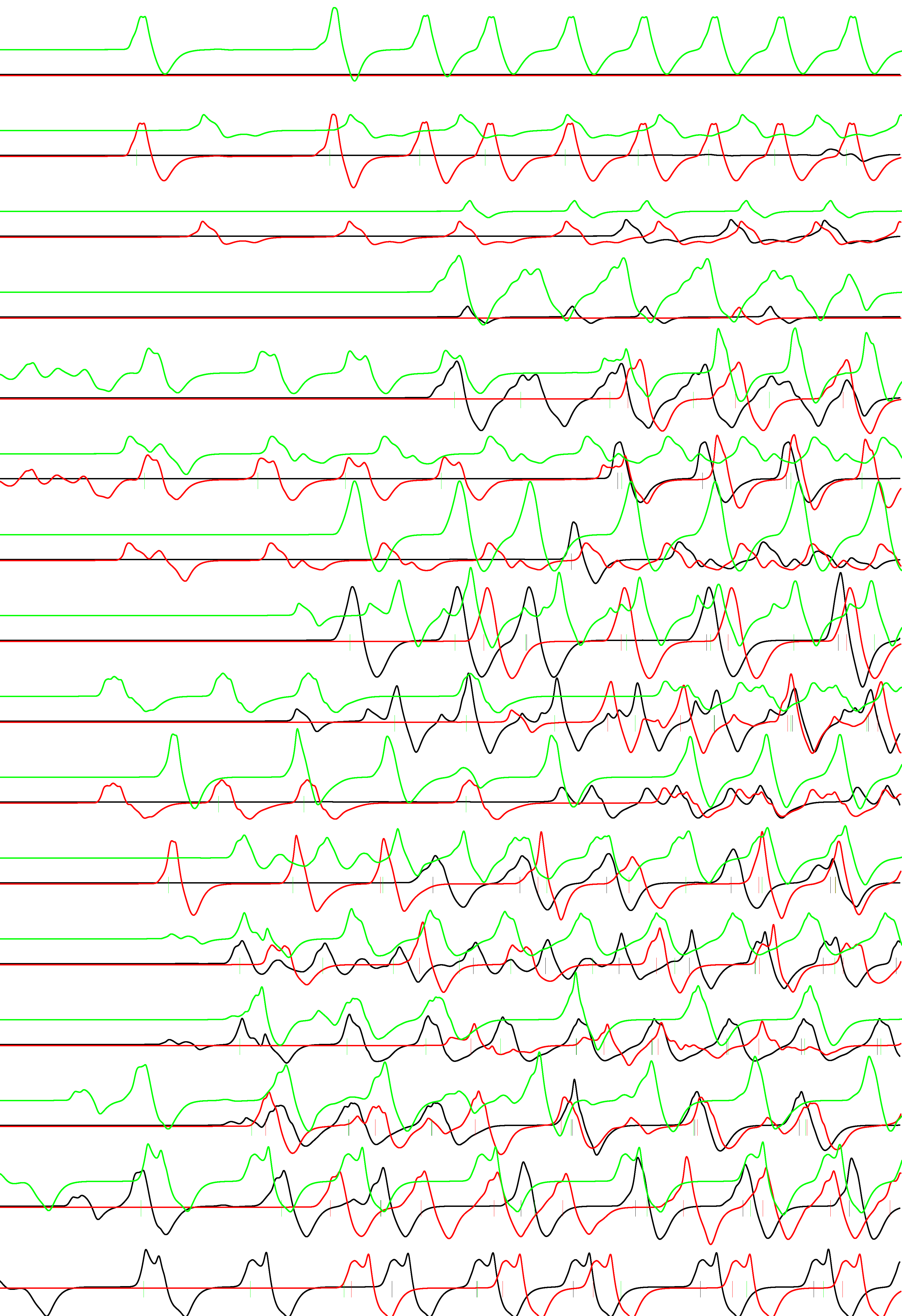}
    \caption{Recording of electrical potential from all electrodes in responses to inputs in response to inputs (01), black  line, (10), red  line, (11), green line, injected as spikes via electrodes $E_x=5$ and $E_y=15$. }
    \label{fig:5_15_allrecordings}
\end{figure}

We adopt the encoding procedure developed by us in~\cite{adamatzky2019computing}. We select two electrodes as inputs  $x$ and $y$. We represent logical {\sc True}, or `1' as an impulse injected in the network via input electrode. For example, if $x=1$ then the site corresponding to $x$ is excited, if $x=0$ the site is not excited. 

Each spike represents logical {\sc True}. The spikes occurring within less than $2 \cdot 10^2$ iterations are seen as occuring simultaneously.We assume that spikes are separated if their occurrences lie more than  $10^3$ iterations apart. An example is shown in Fig.~\ref{fig:5_15_example}.  

\begin{table}[!tbp]
    \begin{tabular}{c|cccccccc}
$E$	&	$x+y$	&	$Sy$	&	$x\oplus y$	&	$Sx$	&	$\overline{x}y$	&	$x\overline{y}$	&	$xy$	&	Total	\\ \hline
0	&	0	&	0	&	0	&	0	&	0	&	0	&	0	&	0	\\
1	&	0	&	0	&	0	&	2	&	0	&	0	&	0	&	2	\\
2	&	0	&	0	&	0	&	0	&	0	&	0	&	0	&	0	\\
3	&	0	&	0	&	0	&	0	&	0	&	0	&	0	&	0	\\
4	&	1	&	0	&	0	&	7	&	1	&	0	&	0	&	9	\\
5	&	0	&	0	&	0	&	2	&	2	&	0	&	0	&	4	\\
6	&	0	&	0	&	0	&	2	&	0	&	0	&	0	&	2	\\
7	&	1	&	0	&	0	&	8	&	2	&	0	&	0	&	11	\\
8	&	1	&	0	&	0	&	6	&	1	&	0	&	0	&	8	\\
9	&	0	&	0	&	0	&	0	&	1	&	0	&	0	&	1	\\
10	&	0	&	1	&	1	&	2	&	0	&	1	&	2	&	7	\\
11	&	0	&	0	&	0	&	4	&	2	&	0	&	0	&	6	\\
12	&	0	&	0	&	0	&	3	&	2	&	0	&	0	&	5	\\
13	&	1	&	5	&	0	&	0	&	0	&	1	&	0	&	7	\\
14	&	2	&	5	&	0	&	1	&	0	&	1	&	0	&	9	\\
15	&	0	&	1	&	0	&	5	&	2	&	0	&	0	&	8	\\
Total	&	6	&	12	&	1	&	42	&	13	&	3	&	2	&	79	\\
    \end{tabular}
    \vspace{0.5cm}
    \caption{Numbers of Boolean gates detected for selected pairs of input electrodes 3 and 13. }
    \label{tab:numbers}
\end{table}

\begin{figure}[!tbp]
    \centering
    \includegraphics[width=0.9\textwidth]{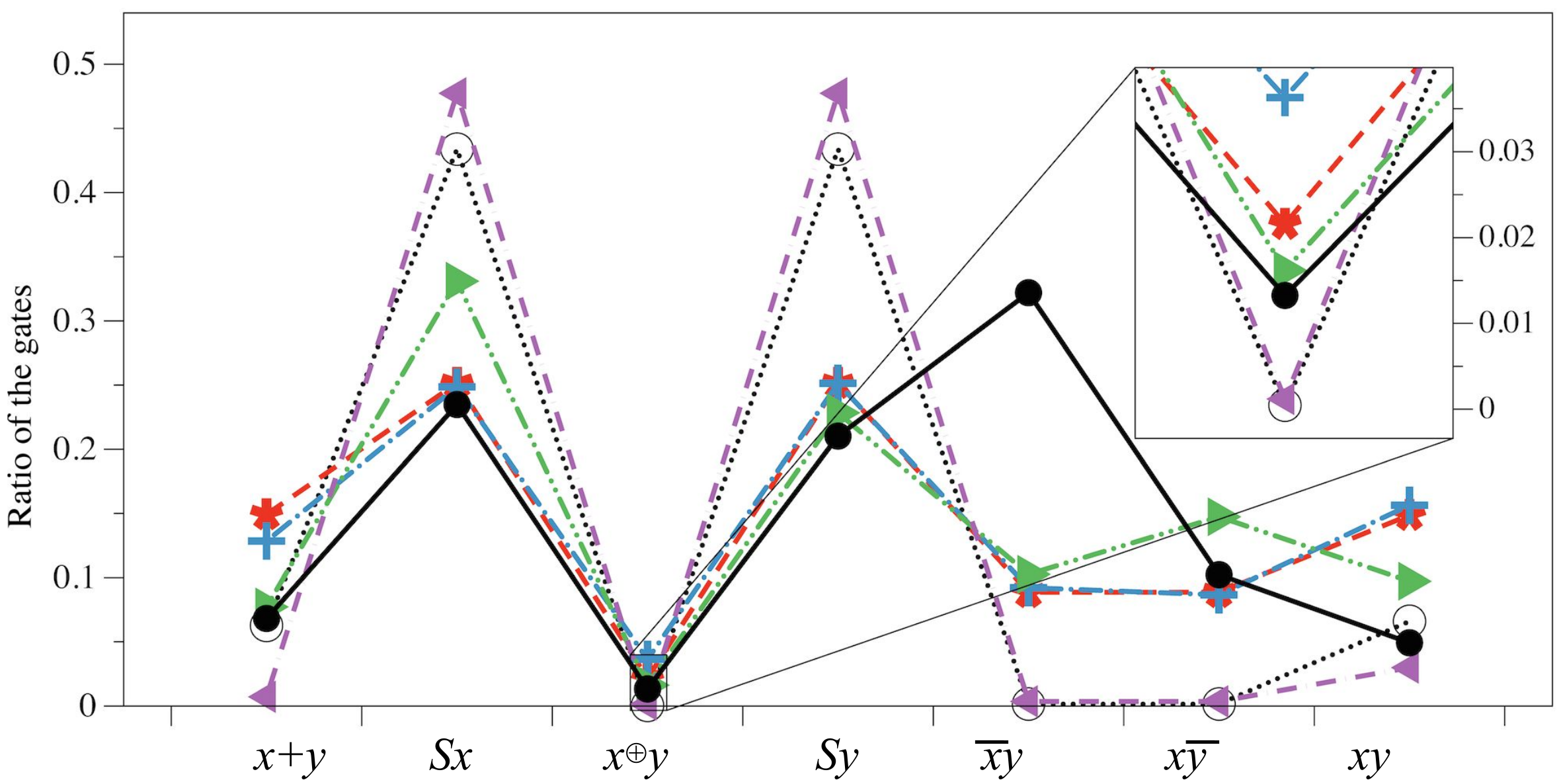}
    \caption{Comparative ratios of Boolean gates discovered in 
    mycelium network analysed in present paper, black disc and solid line;
    slime mould \emph{Physarum polycephalum}~\cite{harding2018discovering}, black circle and dotted line;
    succulent plant~\cite{adamatzky2018computers}, red snowflake and dashed line; 
    single molecule of protein verotoxin~\cite{adamatzky2017computing}, light blue `+' and dash-dot line;
    actin bundles network~\cite{adamatzky2019computing}, green triangle pointing right and dash-dot-dot line;
    actin monomer~\cite{adamatzky2017logical}, magenta triangle pointing left and dashed line. Area of {\sc xor} gate is magnified in the insert. Lines are to guide eye only.}
    \label{fig:gatesDistribution}
\end{figure}

Numbers of Boolean gates detected on the electrodes for selected pair of input electrodes are shown in Tab.~\ref{tab:numbers}. The most frequent gates are  select $x$ and select $y$ gates and occur similar frequencies. 
The {\sc and-not} gates  $\overline{x}y$ and $x\overline{y}$ less common than select gates. The gates $xy$ and $x+y$ are detected with nearly the same frequency with gate $x+y$ being slightly more common. The most rare gate is a logical exclusion  $x\oplus y$. 

The overall distribution of the ratio of gates discovered is shown in Fig.~\ref{fig:gatesDistribution}. The distribution demonstrates frequencies of discoveries of the four-input-one-output logical gates and could be used for characterisation of a computational power of the fungal substrates. This is accompanied by distributions of gates discovered in experimental laboratory reservoir computing with slime mould \emph{Physarum polycephalum}~\cite{harding2018discovering}, succulent plant~\cite{adamatzky2018computers} and numerical modelling experiments on computing with protein verotoxin~\cite{adamatzky2017computing},    actin bundles network~\cite{adamatzky2019computing}, and actin monomer~\cite{adamatzky2017logical}. The distributions of gates discovered in natural systems are alike to each other in the hierarchies of the gates frequencies. Namely, gates selecting one of the inputs are most common, they are followed by {\sc or} gate, then by {\sc not-and} an {\sc and-not} gates. The gate {\sc and} is typically  less frequent. The gate {\sc xor} is a totally rare. 

\subsection{Resistive and capacitive (RC) networks}

\begin{figure}[!tbp]
    \centering
    \subfigure[]{\includegraphics[width=0.99\textwidth]{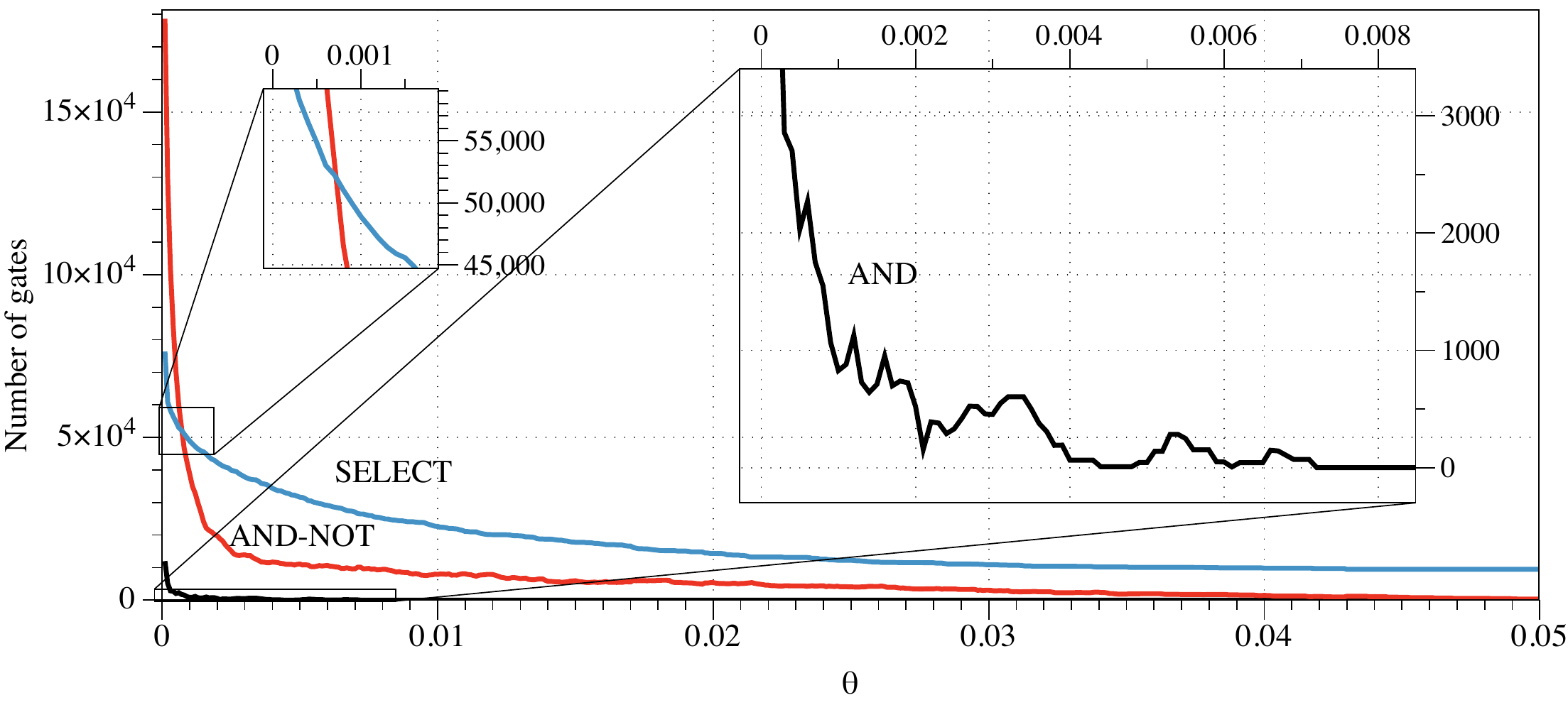}}
    \subfigure[]{\includegraphics[width=0.99\textwidth]{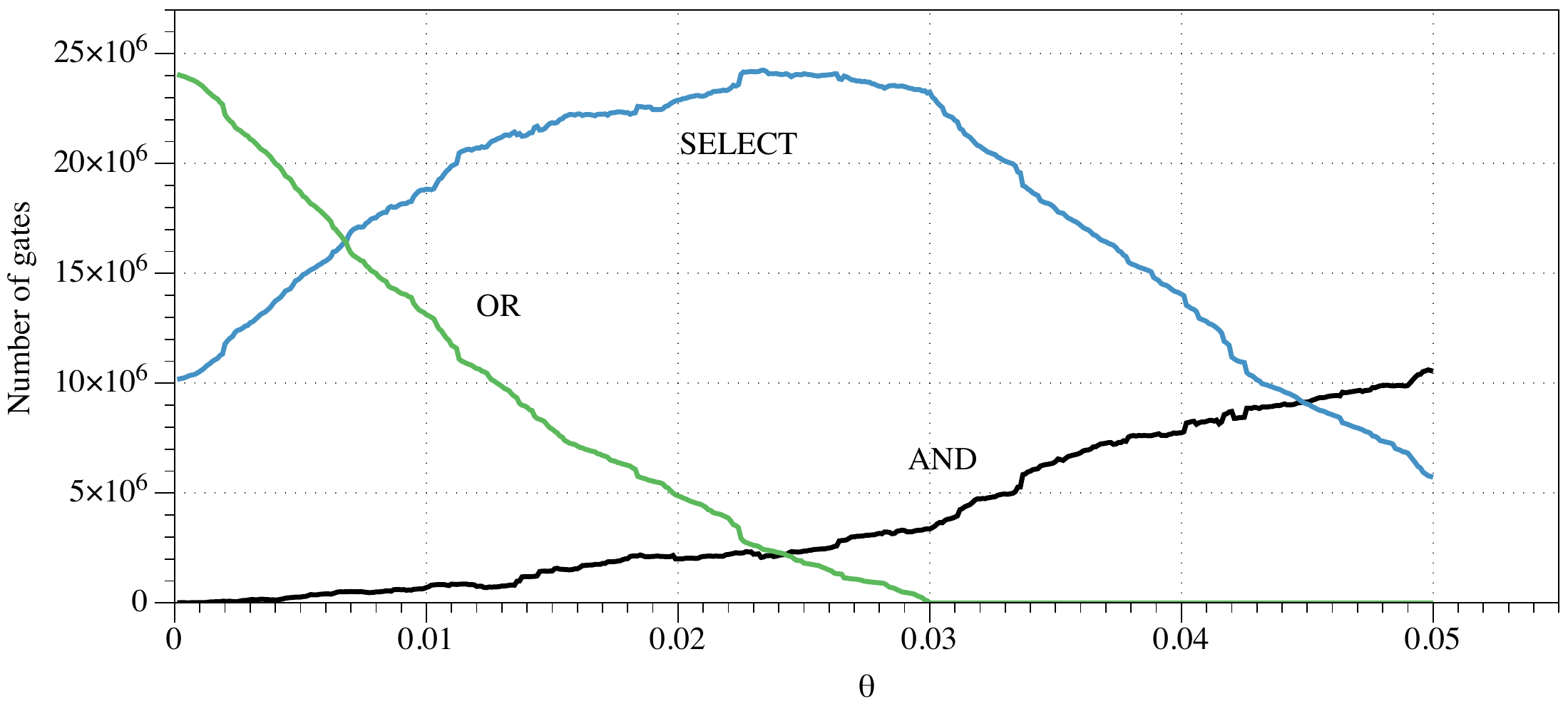}}
    \caption{Occurrences of the gates from the groups {\sc and}, black, {\sc or}, green, {\sc and-not}, red, and {\sc select}, blue, for $\theta \in [0.0001, 0.05]$, with $\theta$ increment 0.0001, in (a)~fungal colony modelled with serial RC networks, (b)~fungal colony modelled with parallel RC networks.}
    \label{fig:gates}
\end{figure}

There are sixteen types of two-input-one-output Boolean gates. The `active' gates, i.e. those where zero inputs evoke a non-zero response could not be realised in the passive electrical model of a fungal colony. They realisable gates are   
{\sc and}, {\sc or}, 
{\sc and-not} ($x$ {\sc and not} $y$ and {\sc not} $x$ {\sc and} $y$), {\sc select} ({\sc select} $x$ and {\sc select} $y$) 
and {\sc xor}. The exclusion gates {\sc xor}  have not been detected in any of the RC models of the fungal colony.  

In the model of serial RC networks, we found gates {\sc and}, {\sc select} and {\sc and-not}; no {\sc or} gates have been found. The number $n$ of the gates discovered decreases by a power law with increase of $\theta$: 
$n_{{\text {\sc and-not}}}=72 \cdot  x^{-0.98}$, $n_{{\text {\sc select}}}=2203 \cdot x^{-0.48}$, $n_{{\text {\sc and}}}=0.02 \cdot x^{-1.6}$. Frequency of {\sc and} gate oscillates, as shown in zoom insert in Fig.~\ref{fig:gates}a, more likely due to its insignificant presence in the samples. The oscillations reach near zero base when $\theta$ exceeds 0.001.

In the model of parallel RC networks we found only gates {\sc and}, {\sc select} and {\sc or}. The number of {\sc or} gates decreases quadratically and becomes nil when $\theta>0.03$. The number of {\sc and} gates increases near linearly, $n_{{\text {\sc and}}}= -1.72 \cdot 10^6 + 2.25 \cdot 10^8 \cdot x$, with increase of $\theta$. The number of {\sc select} gates reaches its maximum at $\theta=0.023$, and then starts to decreases with the further increase of $\theta$: $n_{{\text {\sc select}}}= 9.61 \cdot 10^6 + 1.21 \cdot 10^9 \cdot x - 2.7 \cdot x^2$. 

\subsection{Experimental laboratory mining}

\begin{figure}[!tbp]
    \centering
    \includegraphics[width=\textwidth]{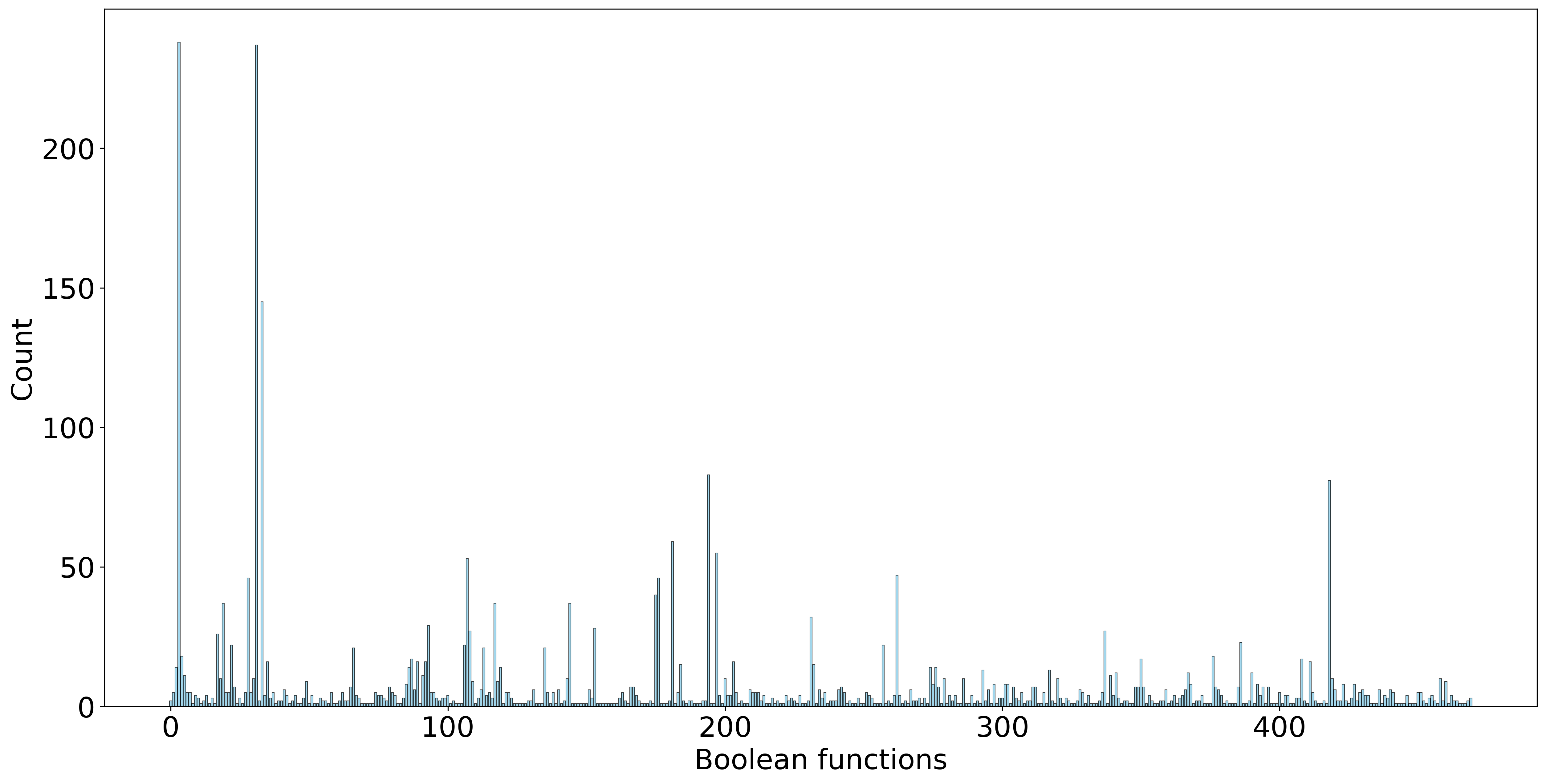}
    \caption{Counts of realised Boolean functions discovered in laboratory experiments. Horizontal axis is a decimal representation of functions. Vertical axis is a number of functions discovered in experiments.}
    \label{fig:Bool_func_count}
\end{figure}

We have discovered total of 3136 4-inputs-1-output Boolean functions. 470 unique functions are detailed in \cite{roberts2021mining}. Figure~\ref{fig:Bool_func_count} shows the Boolean function distribution. The two peak values were logical {\sc False}, $n=238$, and logical {\sc True}, $n=237$. The highest occurring non-trivial gate was $\overline{A}+\overline{B}+\overline{C}+\overline{D}$, $n=145$. The top nine occurring non-trivial Boolean functions are listed in table~\ref{table:1}. The only single gate functions found were for {\sc nand} ($\overline{A}+\overline{B}+\overline{C}+\overline{D}$), $n=145$, {\sc or} ($A+B+C+D$), $n=46$, and {\sc and} ($ABCD$), $n=8$.

\begin{table}[!tbp]
\centering
  \begin{tabular}{ |p{1cm}|p{1cm}|p{8cm}|  }
   \hline
   Count & & Boolean function\\
   \hline
    145 & $F_1$ &$\overline{A}+\overline{B}+\overline{C}+\overline{D}$ ({\sc nand})\\
    83 & $F_2$ & $A\overline{B}+A\overline{C}+A\overline{D}+\overline{A}B+B\overline{C}+B\overline{D}+\overline{A}C+\overline{B}C+C\overline{D}+\overline{A}D+\overline{B}D+\overline{C}D$\\
    81 & $F_3$ & $AC\overline{D}+\overline{A}B\overline{C}+\overline{A}\overline{B}C+\overline{A}\overline{B}D$\\
    59 & $F_4$ & $A\overline{C}+A\overline{D}+\overline{A}C+C\overline{D}+\overline{A}D+\overline{B}D+\overline{C}D$\\
    55 & $F_5$ & $\overline{A}B+C\overline{D}+\overline{A}D$\\
    53 & $F_6$ & $A\overline{B}CD$\\
    47 & $F_7$ & $B\overline{D}+C\overline{D}+\overline{A}D+\overline{B}\overline{C}D$\\
    46 & $F_8$ & $AB\overline{C}\overline{D}$\\
    46 & $F_9$ & $A+B+C+D$ ({\sc or})\\
   \hline
  \end{tabular}
 \label{table:1}
 \vspace{0.5cm}
 \caption{Top nine highest occurring Boolean functions discovered in experimental laboratory mining with a substrate colonised by living mycelium.}
\end{table}

\section{Discussion}

In numerical modelling and experimental laboratory studies we demonstrated that a wide range of Boolean circuits are implemented in a single fungal colony and a substrate colonised by mycelium. In the models where logical functions are implemented with spikes (travelling excitation waves), the {\sc xor} gate is the rarest, {\sc or} and {\sc and} are more common and {\sc and-not} are most common ({\sc select} is a rather trivial gate). The frequency distribution of the gate is generally in line with the distributions of gates discovered in other living substrates. In the resistive and capacitive (RC) network model of a single fungal colony, we discovered {\sc and-not} gate in serial networks, and {\sc or} and {\sc and} in parallel networks. This relatively poor representation of logical functions might be due to the absence of capacitive elements. In contrast to the RC model, sets of logical circuits discovered in laboratory experiments with living mycelium are impressively large~\cite{roberts2021mining}. This is because living mycelium networks are active, i.e. they generate spikes of electrical potential~\cite{adamatzky2018spiking} and spikes of resistance~\cite{adamatzky2021electrical}, capacitive~\cite{beasley2020capacitive} and memfractive properties~\cite{beasley2020fungal}.

\subsection*{Acknowledgement}

This research has received funding from the European Union's Horizon 2020 research and innovation programme FET OPEN ``Challenging current thinking'' under grant agreement No 858132 / project \textit{Fungal Architectures}.\\
(www.fungar.eu).



\end{document}